\documentclass[10pt,preprint,aps,prc,showpacs]{revtex4-1}
\usepackage{graphicx,epstopdf,multirow,amsmath,rotating}

\begin{document}

    
		\author{Deepak Kumar}
    \author{Moumita Maiti\footnote{E-mail: moumifph@iitr.ac.in, moumifph@gmail.com (Reprint author)}}
    \affiliation{Department of Physics, Indian Institute of Technology Roorkree, Roorkee-247667, Uttarakhand, INDIA}
	  \title{Measurement of cross section of the residues from the $^{11}$B-induced reaction on $^{89}$Y and $^{93}$Nb: Production of $^{97}$Ru and $^{101m}$Rh}
		
	  \date{\today}
	
	
	\begin{abstract}
		\begin{description}
\item[Background] 
The heavy-ion induced reactions on intermediate mass targets are complex in nature, even at the low energies. To understand those nuclear reaction phenomena in detail, more experimental studies are required in a wide range of energy.

\item[Purpose] 
Investigation of heavy-ion reactions by measuring production cross sections of the residues produced in the $^{11}$B-induced reactions on $^{89}$Y and $^{93}$Nb at low energies, near and above the barrier, and to check the effectiveness of the different nuclear models to explain them. Further, aim is also to optimize the production parameters of neutron deficient medically relevant $^{97}$Ru and $^{101m}$Rh radioisotopes produced in those reactions, respectively.   
\item[Method] 
The $^{11}$B-beam was allowed to impinge on $^{89}$Y and $^{93}$Nb foils supported by an aluminum (Al) catcher foil, arranged in a stack, in 27.5--58.7 and 30.6--62.3 MeV energy range, respectively. The off-line $\gamma$-spectrometry was carried out after the end of bombardment (EOB) to measure the activity of the radionuclides produced in each foil and cross sections were calculated. Measured cross-sectional data were analyzed in terms of compound and precompound model calculations. 
\item[Results] 
The measured cross sections of $^{97,95}$Ru, $^{96,95,94}$Tc, $^{93m}$Mo, $^{90m}$Y radionuclides produced in the $^{11}$B+$^{89}$Y reaction, and $^{101,100,99}$Pd, $^{101m,100,99m}$Rh, $^{97}$Ru produced in the $^{11}$B+$^{93}$Nb reaction showed good agreement with the model calculations based on the Hauser-Feshbach formulation and exciton model. Unlike theoretical estimation, consistent production of $^{90m}$Y was observed in the $^{11}$B+$^{89}$Y reaction. Substantial preequilibrium contribution was noticed in the 3$n$ reaction channel in both the reactions. 
			
\item[Conclusions] Theoretical estimations confirmed that major production yields are mostly contributed by the compound reaction process. Preequilibrium emissions contributed at the high energy tail of the 3$n$ channel for both the reactions. Moreover, an indirect signature of a direct reaction influence was also observed in the $^{90m}$Y production.
     \end{description}
	\end{abstract}
   
	\pacs{24.10.-i, 24.60.Dr, 25.70.-z, 25.70.Gh }
  \maketitle
     
	
	\section{\label{s1}Introduction}
	The comprehensive study on fusion reactions of tightly and weakly bound stable heavy projectiles with medium and heavy targets has been performed and discussed in the literature for many decades \cite{back14,canto15}. However, strength of reaction processes, such as complete-incomplete fusion, deep inelastic scattering, quasifission, precompound processes near the Coulomb barrier, nucleon transfer reactions etc.,\ observed in the low-energy heavy ion induced reactions have not yet been understood in a full-fledged fashion \cite{parker91,kelly00,dasgupta04,zolnowski78,timmers98,vergani93,sharma15,shrivastava06}. In spite of that, its importance to understand the synthesis of superheavy elements \cite{hinde11,hinde08}, nucleosynthesis process necessary for the star evolution\cite{wallerstein97}, quantum mechanical tunneling phenomenon near sub-barrier energies, fusion hindrance phenomenon at deep sub-barrier energies \cite{jiang05,stefanini08}, behavior of nuclei far from the stability region investigated using radioactive ion beam on the medium/heavy mass targets \cite{rehm98,vinod13,signorini97}, etc.\ fascinate researchers and make it a central topic for nuclear physics research in recent years.
	
	In view of this quest, a systematic study of boron ($^{11}$B) induced reaction on intermediate mass targets: yttrium($^{89}$Y) and niobium($^{93}$Nb), has been reported in the $\sim$2.5--5.5 MeV/nucleon energy range.\ Production of $^{97,95}$Ru, $^{96,95,94}$Tc, $^{93m}$Mo, $^{90m}$Y and $^{101,100,99}$Pd, $^{101m,100,99m}$Rh, $^{97}$Ru radionuclides are observed at $^{89}$Y and $^{93}$Nb targets, respectively. This article contributes to study the nuclear reaction mechanisms involved in the production of various radioisotopes.  It is also important to examine the reaction cross section predictive capability of recent upgrades in modern reaction codes for model calculations. Optimization of the production of neutron deficient medically relevant radionuclides, $^{97}$Ru and $^{101m}$Rh, which are produced in $^{11}$B+$^{89}$Y and $^{11}$B+$^{93}$Nb reactions, respectively, has also been discussed.   

	The $^{97}$Ru radionuclide could be used in both diagnostic and therapeutic processes due to its suitable half-life (2.83 d), low-lying intense $\gamma$-rays: 215.70 keV (85.62\%) and 324.49 keV (10.79\%) energy, and high chemical reactivity.\ It could be produced in the no-carrier-added (NCA) state, which is the prerequisite of clinical applications \cite{maiti09}, from $^{11}$B+$^{89}$Y reaction.\  
	 Initially, $^{97}$Ru was produced mainly by the bombarding high energy proton beams on different targets: rhodium ($^{103}$Rh), technetium ($^{99}$Tc), \cite{srivastava78,lagunas83,zaitseva92}. Besides, $\alpha$-particle or $^3$He induced reactions on natural molybdenum targets were also led to the production of $^{97}$Ru \cite{comar76,omperetto80,pao81}. Recently, heavy ion ($^{7}$Li, $^{11}$B, $^{12}$C) induced reactions on natural Nb and Y metals were also studied by our group to produce NCA $^{97}$Ru \cite{maiti11,maiti15,maiti13,deepak16,dkumar17}.
	
	On the other hand, low-lying high intense $\gamma$-rays of $^{101m}$Rh help in the $in\ vivo$ monitoring by using a scintillation camera, while emission of Auger, Coster-Kronig electrons and X-rays aid in therapeutic applications. The $^{101m}$Rh (4.34 d) decays mainly by electron capture (92.8 \%) to stable $^{101}$Ru by emitting 306.9 keV (81 \%) and 545.1 keV (4.3 \%) $\gamma$-rays and by isomeric transition (IT) (7.2 \%) to $^{101}$Rh (3.3 a), which finally decays to $^{101}$Ru.
	
So far, production of $^{101m}$Rh has been studied using light ion ($p$, $d$, $^{3}$He, $^{4}$He) induced reactions only. $^{101m}$Rh was produced through the decay of its precursors $^{101}$Pd and $^{101}$Ag, which were populated in $p$-induced reactions on the $^{103}$Rh, $^{nat}$Pd, and $^{nat}$Ag targets \cite{scholtz77,lagunas82,lagunas84,ditroi07,hermanne00,sudar02,uddin05}. In the deuteron-induced reactions, production of $^{101m}$Rh using $^{103}$Rh target was measured by Hermanne $et$ $at.$, Detr$\acute{o}$i $et$ $al.$ in different energy  regions \cite{hermanne02,ditroi11,hermanne15}. Direct and cumulative production of $^{101m}$Rh in the d+$^{nat}$Pd reaction was also experimented by Detr$\acute{o}$i $et$ $al.$ \cite{ditroi12}, however the production was significantly low. Apart from these, production of $^{101m}$Rh isomer was reported by Skakun $et$ $al.$ \cite{skakun08} for the $^3$He-induced reaction via $^{101}$Ru($^3$He,$x$)$^{101m}$Rh (cum) and $^{102}$Ru($^3$He,$x$)$^{101m}$Rh reactions.
	 
	Thus we have gone through the current interest to study the heavy ion induced reactions and the production of $^{97}$Ru and $^{101m}$Rh radionuclides in those reactions. The experimental procedure and brief review of the nuclear model calculations are described in Secs. \ref{s2} and \ref{s3}, respectively. Section \ref{s4} discusses the results and Sec. \ref{s5} finally concludes the report.
	
    \section{\label{s2}Experiment}
    
\begin{figure} 
	  \includegraphics[width=90mm]{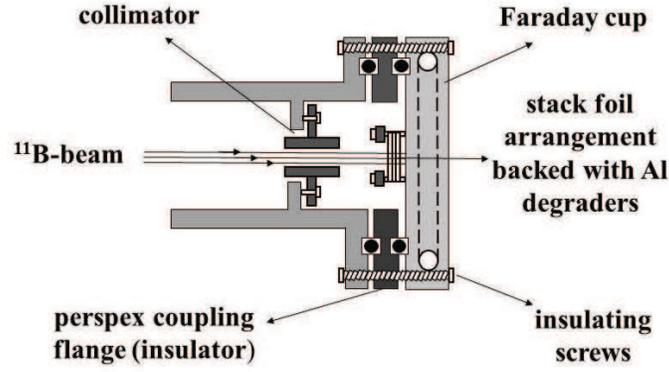}
	  \caption{\label{fig1} A schematic diagram of a typical stack foil activation setup.}
       \end{figure}

     The experiment was performed at the BARC-TIFR Pelletron facility, Mumbai, India. Stacked foil activation technique was used to explore $^{89}$Y+$^{11}$B and $^{93}$Nb+$^{11}$B reactions. A stack of targets was prepared by placing 2-3 target foils each one of which was backed by an aluminum (Al) catcher foil so that target and backing appear in an alternative fashion. A typical experimental setup to irradiate the stack foil arrangement is shown in Fig \ref{fig1}. Each target stack was bombarded by the $^{11}$B$^{5+}$ beam for a stipulated time. Beam current was kept almost constant during the experiment. The total charge passed through the target was measured by an electron suppressed Faraday cup placed at the back of the target assembly during the beam-on period. The time of irradiation was decided by the projectile flux and the half-lives of the product radioisotopes. A total of 4-5 such stack was irradiated individually for each target-projectile combination varying the incident energy of $^{11}$B with a slight overlap between them. 
		
 Spectroscopically pure (99.99\%) natural yttrium and niobium foils were procured from the Alfa Aesar. Self-supporting thin foils of those metals were made by uniform rolling in a machine. Thicknesses of the thin Y- and Nb-foils were between 0.84-2.9 mg/cm$^2$ and 1.2-2.1 mg/cm$^2$, respectively. The Al-backing foils had thickness between 1.5-2 mg/cm$^2$. 
Annular aluminum holders with an inner and outer diameter 12 and 20 mm, respectively, were used to mount the foils. The large area of Al foil ensures the complete collection of recoiled the residues, if recoiled any, in the beam direction. The Al-foils also served the purpose of an energy damper so that suitable energy separation between consecutive target foils could be achieved. Energy degradation of the projectile in each foil was estimated by Stopping and Range of Ions in Matter (SRIM) code \cite{ziegler10}. The $^{11}$B energy at a target is typically the average of the incident and outgoing beam energy.  
	
	After the end of bombardment (EOB), the residual radionuclides produced in each target foil (Y and Nb) were identified and quantified with the help of $\gamma$-ray spectrometry using a broad energy germanium (BEGe) based detector attached with a PC operating with the GENIE-2K software. The detector was calibrated by using standard sources, $^{152}$Eu (13.506 a), $^{137}$Cs (30.08 a), $^{60}$Co (5.27 a), $^{133}$Ba (10.51 a), of known activity. The energy resolution of the detector was $\le $ 2.0 keV at 1332 keV energy. 
		
	The activity of the residuals was measured in a fixed geometry over a longer period of time in the regular intervals to follow their decay profile. The activity of the residuals at the EOB was measured from the background-subtracted peak area count rate (counts/sec). Production cross section of the residues at each incident energy was calculated from the activation formula. The detail description of the activity and cross-section measurement is available in our previous reports \cite{deepak16,ms11,m11}. Nuclear spectroscopic data of the residual radionuclides are listed in Table \ref{tab1} \cite{nndc}.
	
	The error introduced in the cross section measurement was mainly due to non-uniformity of samples and in measuring its thickness ($\sim$5\%), fluctuation of beam current ($\sim$5\%), efficiency calibration of the detector ($\sim$2\%). Some other sources such as the branching intensity of characteristic $\gamma$-rays, counting statistics, beam energy degradation while traversing through the successive target foils (straggling effects) etc.\ are also responsible; however, these were negligible in this case \cite{wilke76,kemmer80}. The total uncertainty associated with the cross section measurement was determined considering all those factors and the data was presented in this article up to 95\% confidence level.

	\section{\label{s3}Analysis of measured data}
	
    \subsubsection{PACE4} 
		Compound reaction contribution in the produced residues at different bombarding energies in the $^{11}$B+$^{89}$Y and $^{11}$B+$^{93}$Nb reactions was extracted with the help of a statistical model code \textsc{PACE4} \cite{gavron80}, build on the Hauser-Feshbach (HF) formulation where angular momentum coupling is considered in each stage of de-excitation of an excited compound nucleus. The compound nucleus cross section has the form 
		\begin{equation}
		\label{eq1}
		\sigma^{HF}_{\alpha\beta}(E_\alpha)=\sum_{J,\Pi}\bigg[\pi\lambdabar^2\frac{2J+1}{(2s+1)(2I+1)}\sum_{j,l}T_{\alpha jl}(E_{\alpha},J,\Pi)\bigg]\bigg[\frac{\sum_{j',l'}  T_{\beta j'l'}(E_\beta,J,\Pi)}{\sum_{\gamma,j'',l''}  T_{\gamma j''l''}(E_\gamma,J,\Pi)}\bigg]
		\end{equation} 
		where $\lambdabar$ ($=\lambda/2\pi$) is the wavelength of incident projectile; $s$, $I$ and $J$ represent projectile, target and compound nucleus spin, respectively, and $l$ is orbital angular momentum of the projectile. $T_{\alpha jl}(E_{\alpha},J,\Pi)$ stands for the transmission coefficient having channel energy $E_\alpha$ and orbital angular momentum $l$, which together with particle spin $s$ couples to the channel angular momentum $j$ used to select in target nucleus spin $I$ populated for a given compound nucleus spin $J$ and parity $\Pi$. Similarly, $T_{\beta j'l'}(E_\beta,J,\Pi)$ and $T_{\gamma j''l''}(E_\gamma,J,\Pi)$ indicate transmission coefficients for $\beta$ and $\gamma$ channels with emitting energies $E_{\beta}$, $E_{\gamma}$ and orbital angular momenta $l'$, $l''$ couple with emitting particle spins gives channel spins $j'$, $j''$, respectively. The quantity in the first square bracket indicates the compound nucleus formation/fusion cross section in a state of total spin ($J$) and parity ($\Pi$) associated to the incident channel $\alpha$ which reduces to a simple form for spin less target ($I=0$),
			\begin{equation}
			\label{eq2}
			\sigma_f = \pi\lambdabar^2 \sum_l (2l+1) T_l 
			\end{equation}
The quantity in the second square bracket represents the decay probability of the compound nucleus in the channel $\beta$ having ejectile particle energy $E_\beta$.		 
			  
\textsc{PACE4} uses Bass model potential \cite{bass77} for the calculation of transmission coefficient, although it is not well suited near/below the barrier as well as for very heavy ion projectiles. Fission is considered as a decay mode, and modified rotating liquid drop fission barrier (by A. J. Sierk) is selected. Evaporation of seven light particles/nuclei in the order n, p, $\alpha$, d, t, $^3$He, $^6$Li is considered whose transmission coefficients are determined by optical model potential, where optical model parameter is taken from Ref \cite{perey76}. Gilbert Cameron (GC) nuclear level density parameter and GC spin cutoff parameter are adopted for the calculation. Little a ratio, a$_f$/a$_\gamma$, is taken as unity. In order to simulate $\gamma$ multiplicity and corresponding energy, a non-statistical yrast cascade gamma decay chain is artificially included. 
	   
    \subsubsection{EMPIRE3.2} EMPIRE considers all three major nuclear reaction formalisms - Direct (DIR), Preequilibrium (PEQ) and Compound (EQ). DIR processes are estimated either by coupled channels approach or distorted wave Born approximation (DWBA) \cite{raynal,raynal72}. There are various PEQ phenomenological and quantum models for light-ion induced reactions. However, PEQ emissions in the heavy-ion induced reactions are not well tested for phenomenological hybrid Monte Carlo simulation as well as quantum mechanical multi-step direct (MSD) and multi-step compound (MSC) models, and therefore ignored by the code. PEQ emission for heavy ion projectiles is calculated using the exciton model, which has the capability to treat the cluster emission built on the Iwamotto Harada model \cite{iwamoto82}. The differential cross section of the PEQ emission has a form
     \begin{equation}
     \label{eq3}
     \frac{d\sigma_{\alpha,\beta}}{d\epsilon_\beta} = \sigma^{CF}(\epsilon_\alpha) \sum_n W_\beta(E,n,\epsilon_\beta) \tau(n),
     \end{equation}
     where $\sigma^{CF}(\epsilon_\alpha)$ is the composite nucleus formation cross section and defined as $\sigma^{CF}(\epsilon_\alpha) = \sigma_{reac}(\epsilon_\alpha) - \sigma_{dir}(\epsilon_\alpha)$, a difference of reaction and direct reaction cross section. $\tau(n)$ is life-time of $n$ exciton configuration.
     $W_\beta(E,n,\epsilon_\beta)$ represents the emission rate of a cluster $\beta$ with energy $\epsilon_\beta$, spin $s_\beta$  and reduced mass $\mu_\beta$ from a state with $n\ (=p+h)$ excitons and can be given as, 
     \begin{equation}
      \label{eq4}
      W_\beta(E,n,\epsilon_\beta) = \frac{2s_\beta + 1}{\pi^2 \hbar^3} \mu_\beta \epsilon_\beta \sigma^{inv}_\beta(\epsilon_\beta) \bigg[\frac{\sum_{\beta}F^\beta_{lm}(\epsilon_\beta)Q_\beta^{lm}(p,h)\omega_{res}(p-l,h,E-\epsilon_\beta-B_\beta)}{\omega_{CN}(p,h,E)}\bigg]
      \end{equation} 
      where $\sigma^{inv}_\beta(\epsilon_\beta)$ is the inverse reaction cross section; $\omega(p,h,E)$ is particle-hole state density, calculated by Williams formula \cite{williams71}. $Q_\beta^{lm}(p,h)$ is a factor accounting for the probability of the outgoing cluster $\beta$ being formed with $l$ particles situated above and $m$ below the Fermi surface ($\beta = l + m$) and factor $F^\beta_{lm}(\epsilon_\beta)$ denotes formation probability of the cluster $\beta$ as a function of its energy. 
      
     EQ processes are calculated from the Hauser-Feshbach model, including width fluctuations and the optical model for fission. However, heavy ion fusion cross section is estimated with the help of a simplified coupled channel model (CCFUS) \cite{dasso87}. In CCFUS, collision between two nuclei is considered in the presence of coupling of the relative motion $\vec{r}$ of the projectile to a nuclear collective motion $\zeta$. The Hamiltonian of such system is written as \cite{hagino12}
    \begin{equation}
    \label{eq5}
    H(r,\zeta) = -\frac{\hbar^{2}\nabla^2}{2\mu}+ V(r) + H^0(\zeta) + H^\prime(r,\zeta)
    \end{equation}
     where $\mu$ is the reduced mass of the system and $V(r)$ is the sum of Coulomb and nuclear Woods-Saxon potential. $H^0(\zeta)$ and $H^\prime(r,\zeta)$ stand for the intrinsic and coupling Hamiltonian, respectively. The coupled-channels equations for the radial part of the total wave function can be written from such Hamiltonian as
		 \begin{equation}
     \label{eq6}
     \bigg(-\frac{\hbar^{2}}{2\mu}\frac{d^2}{dr^2} + \frac{J(J+1)\hbar^{2}}{2\mu r^2} + V(r) - E + \epsilon_n \bigg)u^J_n(r) = - \sum_{n'} H^\prime_{nn'}(r)u^J_{n'}(r)
     \end{equation}
     where $n$ represents a n$^{th}$ quantum state and $\epsilon_n$ is the energy eigenvalue of the intrinsic Hamiltonian. $J$ is the total angular momentum, the sum of relative orbital angular momentum $l$ and the angular momentum of intrinsic motion $I$. $H^\prime_{nn'}$ = $\big<n\big|H^\prime\big|n'\big>$ is the coupling matrix element. Diagonalizing this matrix element at the barrier, the matching wave functions can be uncoupled roughly. The total transmission probability is obtained by summing over the distribution of transmission probabilities for the eigenbarriers, with weight given by the overlap of the initial state with eigenchannels,
      \begin{equation}
      \label{eq7}
      T_J(E)  = \sum_{I} |P^J_{II'}(E)|^2 
      \end{equation}
      where $I'$ is the angular momentum of collective motion in the entrance channel. Thus the fusion cross section can be calculated by using Eq.\ \ref{eq2} except that the transmission probability is now affected by each reaction channel.
      
		
		In the present calculation, the value of the mean free path parameter for the exciton model is chosen as 1.5 and the equilibrium exciton number varies between 10--14 for 25--65 MeV energy range of the projectile. The complete fusion cross section is calculated up to 21--46 $l_{max}$ value for 25--65 MeV energy range using Christensen-Winther potential \cite{christensen76}. Various nuclear level density options: Gilbert Cameron (GC) level density with Ignatyuk systematic, Generalized Superfluid Model (GSM), and Enhanced Generalized Superfluid Model (EGSM) have been used.

	In all the three level density model, Ignatyuk level density parameter, which consists of two independent parameters: asymptotic value of the level density parameter ($\tilde{a}$) and shell effect damping parameter ($\gamma_s$), is used. In GC model, Ignatyuk systematic is used in which $\tilde{a}$ = 0.154A + 6.3$\times$10$^{-5}$A$^{2}$ and $\gamma_s$ = -0.054. GSM and EGSM use $\tilde{a}$ = $\alpha$A + $\beta$A$^{2/3}$ and $\gamma_s$ = $\gamma_0$A$^{1/3}$ where $\alpha$ = 0.103, $\beta$ = -0.105, $\gamma_0$ = 0.375 are used for GSM and $\alpha$ = 0.0748, $\beta$ = 0.0, $\gamma_0$ = 0.5609 are used for EGSM. 
		
	 
	\section{\label{s4}Results and discussion}
\begin{figure}
		\includegraphics[width=86mm]{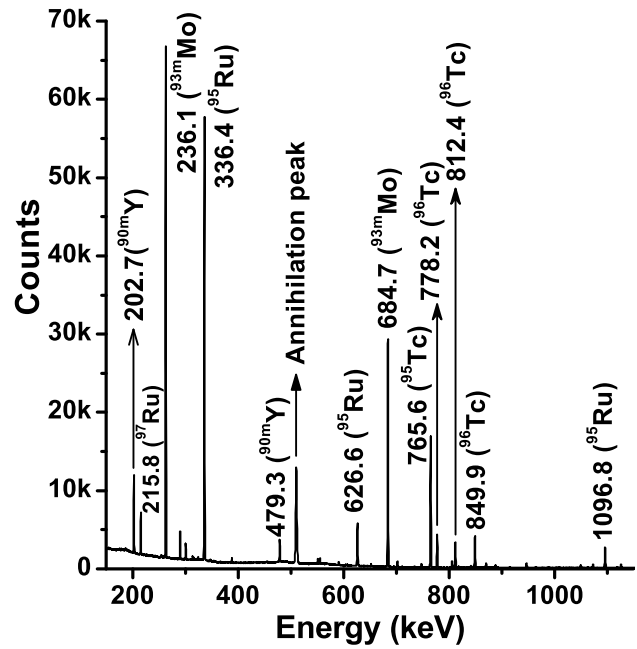}
		\caption{\label{fig2} A $\gamma$-ray spectrum of the $^{11}$B+$^{89}$Y reaction at 50.8 MeV energy after 38.3 minute of EOB.}
		\end{figure}

\begin{figure}
    	\includegraphics[width=86mm]{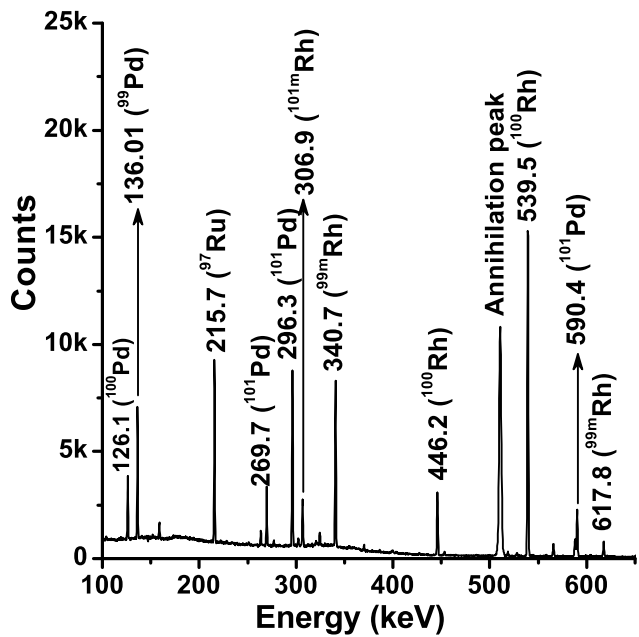}
    	\caption{\label{fig3} A $\gamma$-ray spectrum of the $^{11}$B+$^{93}$Nb reaction at 51 MeV energy after 37.9 minute of EOB.}
    \end{figure}

\begin{figure} 
		\includegraphics[width=86mm]{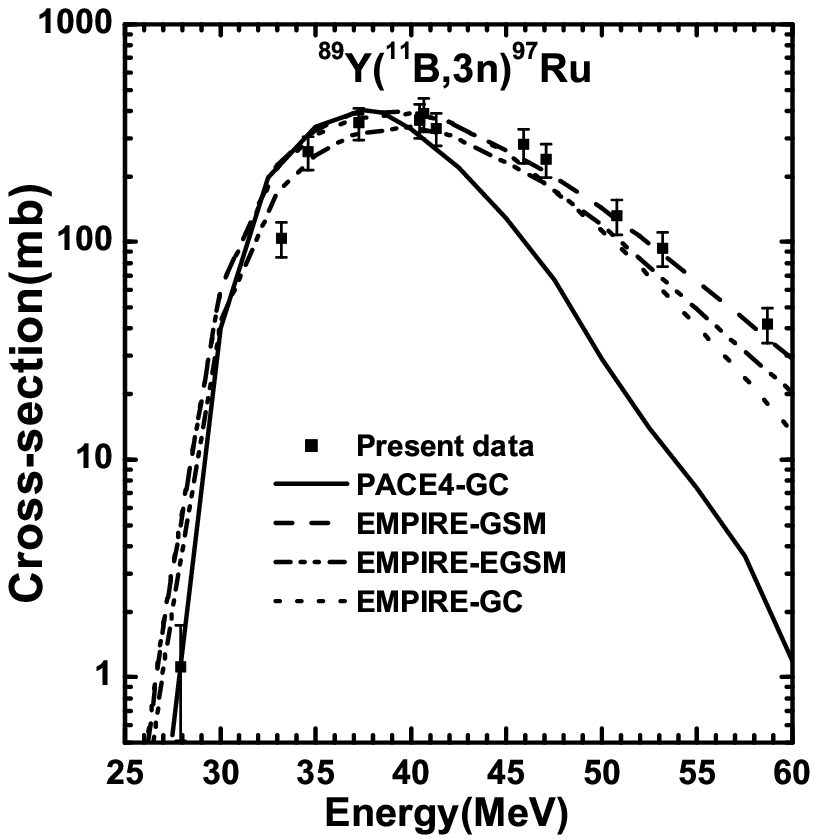}
		\caption{\label{fig4} Comparison of measured excitation function (symbols) of $^{97}$Ru from the $^{11}$B+$^{89}$Y reaction and those computed from the theory (curves) using EMPIRE and PACE4}
		\end{figure}

\begin{figure}
		\includegraphics[width=86mm]{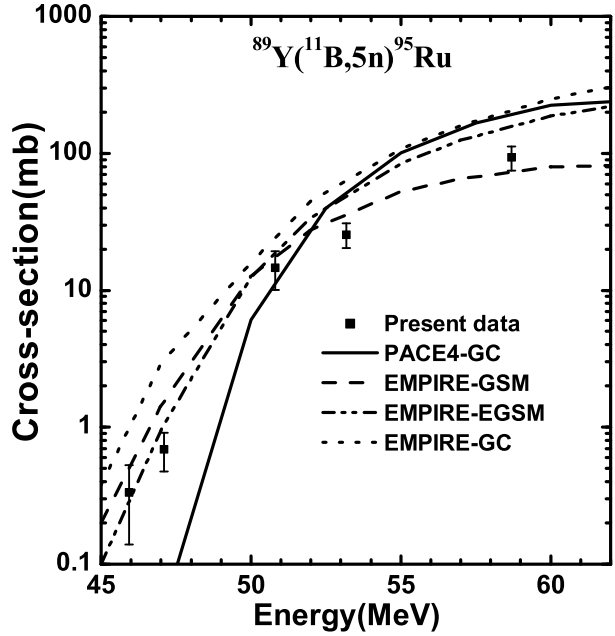}
		\caption{\label{fig5} Comparison of measured (symbols) and computed (curves) excitation functions for production of $^{95}$Ru.}
	\end{figure}

\begin{figure} 
		\includegraphics[width=86mm]{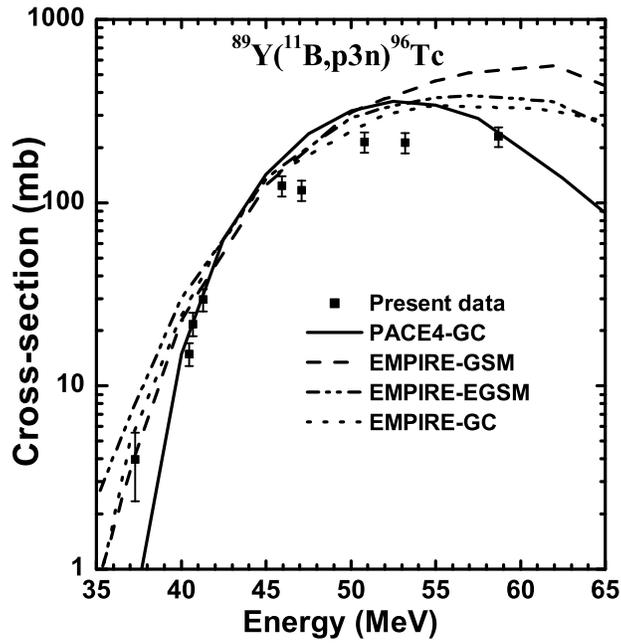}
		\caption{\label{fig6} Same as Fig. \ref{fig5} for $^{96}$Tc.}
		\end{figure}

\begin{figure}
		\includegraphics[width=86mm]{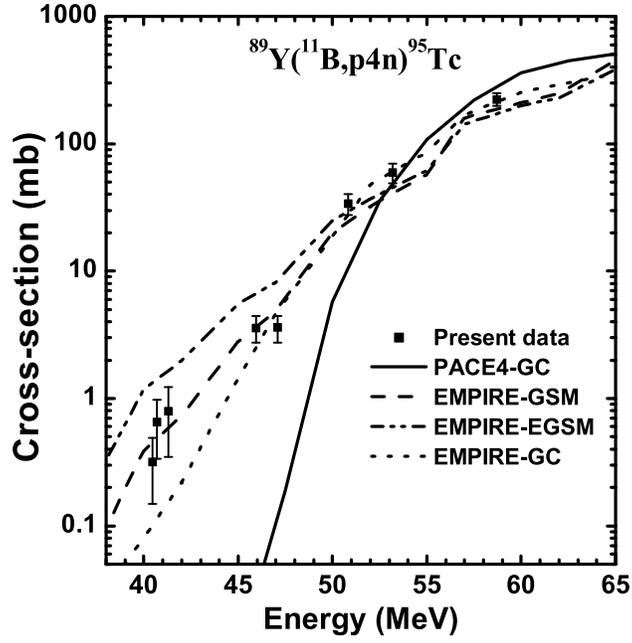}
		\caption{\label{fig7} Same as Fig. \ref{fig5} for $^{95}$Tc.}
	\end{figure}

\begin{figure} [t]
	 	\includegraphics[width=86mm]{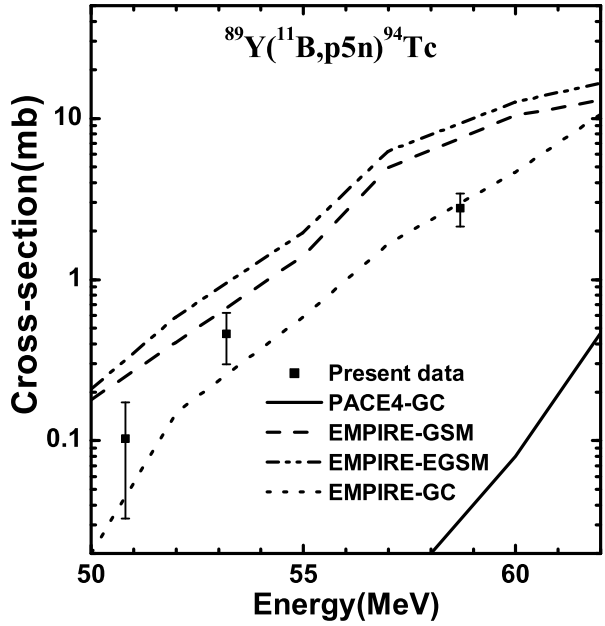}
	 	\caption{\label{fig8} Same as Fig. \ref{fig5} for $^{94}$Tc.}
		\end{figure}

\begin{figure}
	 	\includegraphics[width=86mm]{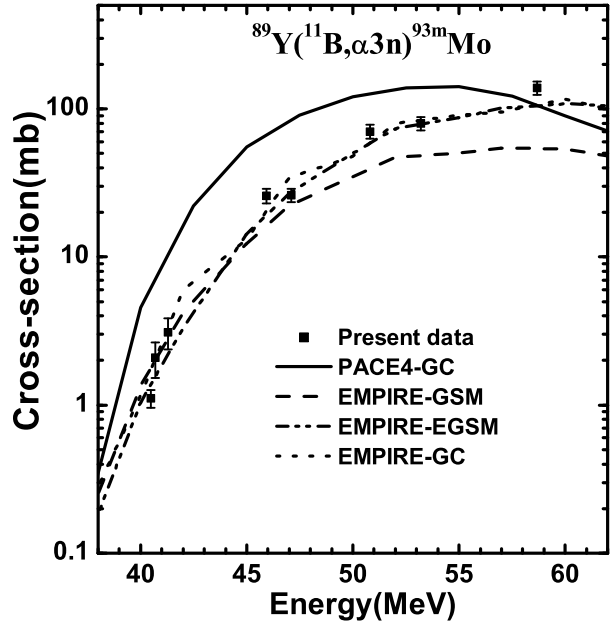}
	 	\caption{\label{fig9} Same as Fig. \ref{fig5} for $^{93m}$Mo.}
	 \end{figure}

\begin{figure} 
	 	\includegraphics[width=86mm]{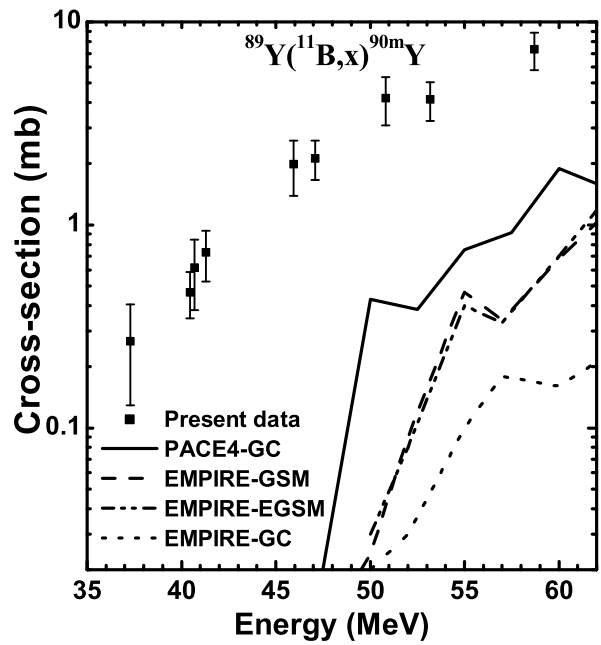}
	 	\caption{\label{fig10} Same as Fig. \ref{fig5} for $^{90m}$Y.}
		\end{figure}

\begin{figure}
	 	\includegraphics[width=86mm]{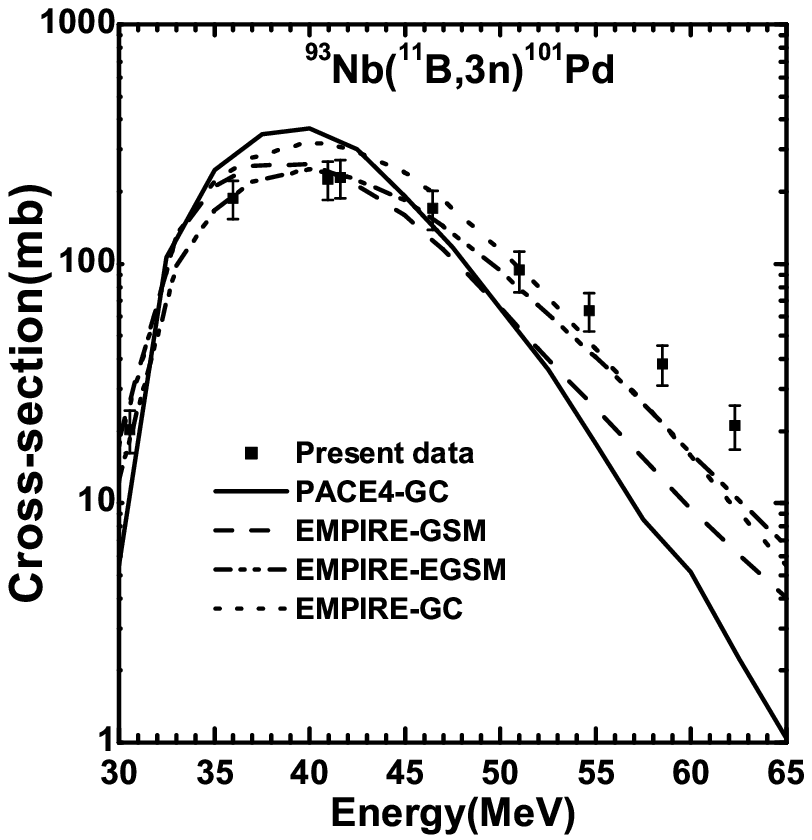}
	 	\caption{\label{fig11} Comparison of measured excitation function (symbols) of $^{101}$Pd from the $^{11}$B+$^{93}$Nb reaction and those computed from theory (curves) using EMPIRE and PACE4}
	 \end{figure}

	 A methodical investigations of radioisotopes produced in the $^{11}$B-induced reaction on $^{89}$Y and $^{93}$Nb targets were carried out for each target foils at various incident energies. Two typical $\gamma$-ray spectra collected 38.3 and 37.9 minutes after the EOB from the $^{11}$B+$^{89}$Y and $^{11}$B+$^{93}$Nb reactions at 50.8 and 51 MeV incident energies are shown in Figs \ref{fig2} and \ref{fig3}, respectively. Production of $^{97,95}$Ru, $^{96,95,94}$Tc, $^{93m}$Mo, $^{90m}$Y residues in the $^{89}$Y; and $^{101,100,99}$Pd, $^{101m,100,99m}$Rh, $^{97}$Ru radionuclides in the $^{93}$Nb target matrix are confirmed and are presented in Figs \ref{fig4}--\ref{fig10}, and Figs \ref{fig11}--\ref{fig18}, respectively. Measured data points are indicated by symbols with an uncertainty and the theoretical estimations are shown by curves.
	
	The measured cross sections are compared with the theoretical model calculations from \textsc{PACE4} and \textsc{EMPIRE3.2} utilizing different nuclear level density options. A list of the radionuclides and the possible contributing reactions are listed in Table \ref{tab1} \cite{nndc}. Experimental cross section data are tabulated in Table \ref{tab2} and Table \ref{tab3} for the $^{11}$B + $^{89}$Y and $^{11}$B + $^{93}$Nb reactions, respectively. 
	
	\subsection{Radionuclides produced in $^{11}$B+$^{89}$Y }
	\subsubsection{$^{97,95}$Ru}
	The measured excitation function of $^{97}$Ru, presented in the Fig \ref{fig4}, is compared with  the   \textsc{PACE} and \textsc{EMPIRE} calculation. Experimental data are well reproduced by both the model codes in the lower energy region ($\sim$28--42 MeV), however, \textsc{PACE} estimation sharply decreases compared to the measured data at higher energies and becomes about 20 times smaller than the experimental cross section at 58.7 MeV energy. The experimental results are in good agreement with those computed from the \textsc{EMPIRE} at the high energy region. It certainly indicates the emission of PEQ neutrons, along with evaporated neutrons in the 3$n$ reaction channel, as observed in various heavy-ion induced reactions \cite{vergani93,sharma15,deepak16}. Reproduction of the data describes the necessity of the PEQ exciton model used in \textsc{EMPIRE} for heavy projectiles. Although all the level densities show a good agreement between each other, GSM level density reproduces the measured data more accurately in comparison to the EGSM and GC level densities at high energy tail of the excitation function. 
	
	Figure \ref{fig5} shows a comparison between the measured cross sections of $^{95}$Ru in the 45--60 MeV energy range and the theoretical evaluation. Experimental cross sections are best explained by \textsc{EMPIRE} with GSM level density throughout the energy range, while other two calculations with EGSM and GC level density overestimate the measured data in the high energy range ($\sim$52--60 MeV). On the other hand, \textsc{PACE} result underpredicts the experimental data below 52 MeV and overpredicts above it. Independent productions of $^{97,95}$Ru are measured due to the absence of any precursor.
	
	\subsubsection{$^{96,95,94}$Tc}
   	Out of the two isotopes, $^{96, 96m}$Tc, the ground state of the long-lived $^{96}$Tc (4.28 d) was identified. Direct identification of the $^{96m}$Tc (51.5 m) radionuclide is hardly possible due to its single low intense $\gamma$-ray (778.22 keV, 1.9\% intensity) common with the high intensity (99.76\%) 778.22 keV peak of $^{96}$Tc. \textsc{EMPIRE} predicts the maximum 6\% production of $^{96m}$Tc compared to $^{96}$Tc production. It is expected that activity of $^{96}$Tc must increase in 5--6 h cooling time due to the decay of $^{96m}$Tc (IT--98\%), if produced in the target matrix, into $^{96g}$Tc radionuclide, however,  no such sign was observed. 
Experimental and theoretical excitation functions of the $^{96}$Tc is shown in Fig \ref{fig6}. Both \textsc{PACE4} and \textsc{EMPIRE} data reproduce the experimental results satisfactorily in the lower energy region, but overestimate in the high energy region.

	In the Fig \ref{fig7}, experimental cross sections of the short-lived $^{95}$Tc (20.0 h) are shown along with the  theoretical excitation functions. The measured data are satisfactorily reproduced by the \textsc{EMPIRE} and \textsc{PACE4} in the higher energy region, while \textsc{PACE} underpredicts them in the lower energy range. \textsc{EMPIRE} calculation corresponding to EGSM and GC has shown the best reproduction of measured data throughout the estimated energy range. Cumulative production of $^{95}$Tc may be possible from $^{95}$Ru radionucides via electron-capture ($\epsilon$-decay) process. However, trace of $^{95m}$Tc was not observed in this experiment.  
	
	The $^{94}$Tc radionuclides may be produced in both meta-stable state $^{94m}$Tc (52 m) and ground state $^{94}$Tc (4.88 h). Due to the common $\gamma$-ray peak at 871.09 keV of $^{94m}$Tc and $^{94}$Tc radionuclides, it is not possible to identify $^{94m}$Tc radionuclides directly. However \textsc{EMPIRE} predicts significantly low isomeric $^{94m}$Tc cross section compared to $^{94}$Tc. The measured excitation function of $^{94}$Tc, shown in Fig \ref{fig8}, is well reproduced by \textsc{EMPIRE} with GC level density while another two level densities overestimate the measured data, particularly at the higher energy region. \textsc{PACE} is unable to explain the experimental observation throughout the observed energy region.    
		
    \subsubsection{ $^{93m}$Mo}
    The excitation function of $^{93m}$Mo is plotted in Fig \ref{fig9} to compare with the \textsc{PACE} and \textsc{EMPIRE} calculations. Experimental data are well reproduced by the \textsc{EMPIRE} with the EGSM level density option throughout the energy range considered. Calculations of \textsc{EMPIRE} using other two level densities are found satisfactory below 47 MeV projectile energy and underpredict the data above it. As the code \textsc{PACE} calculates the cross section of $^{93}$Mo, the sum of isomeric and ground state, it overpredicts the experimental data throughout (up to 55 MeV) the energy.
		
   \subsubsection{$^{90m}$Y}
    A consistent production of $^{90m}$Y radionuclide observed in the $^{11}$B+$^{89}$Y reaction is shown in the Fig \ref{fig10}. The \textsc{PACE4} estimation is for $^{90}$Y, hence it overpredicts the \textsc{EMPIRE} calculation which is for $^{90m}$Y. The trend of the measured excitation function is similar to those obtained from the theoretical cross sections, however, theoretical estimation is almost 10 times lower than the experimental results. Perhaps production of $^{90m}$Y is an indication towards the neutron transfer from the projectile to the yttrium target. 
		
		\subsection{Radionuclides produced in $^{11}$B+$^{93}$Nb}
     \subsubsection{ $^{101,100,99}$Pd}  
       A comparison between the measured and theoretical excitation functions of $^{101}$Pd is shown in Fig \ref{fig11}. It is observed that \textsc{PACE} satisfactory explains the measured data below 47 MeV incident energy and falls considerably beyond it. Overall, experimental observation is well reproduced by \textsc{EMPIRE} with EGSM level density within experimental uncertainties, however small deviation is observed above 52 MeV impinging energy. Similar to that of $3n$ channel reaction in $^{11}$B+$^{89}$Y, precompound emission of neutrons is observed over the compound reaction mechanism at the high energy tail of the excitation function. This direct production of $^{101}$Pd is contributed by both precompound and compound nuclear processes.  
	
\begin{figure} 
		\includegraphics[width=86mm]{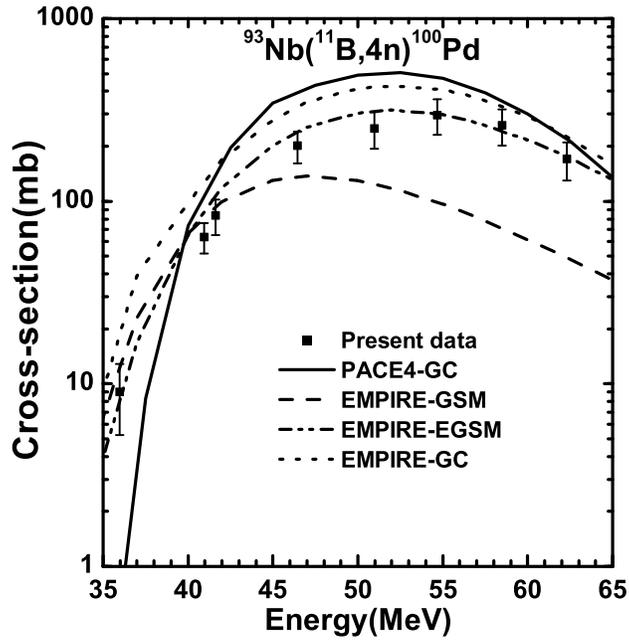}
		\caption{\label{fig12} Same as Fig. \ref{fig11} for $^{100}$Pd.}
		\end{figure}

\begin{figure}
		\includegraphics[width=86mm]{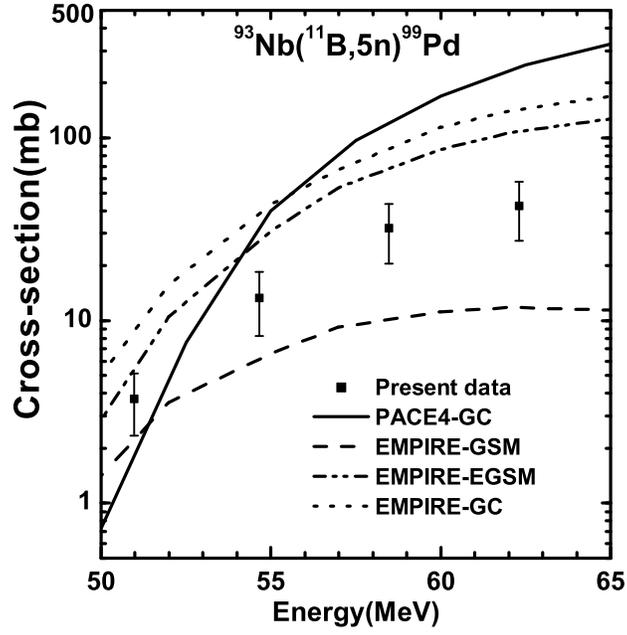}
		\caption{\label{fig13} Same as Fig. \ref{fig11} for $^{99}$Pd.}
	\end{figure}

\begin{figure}
		\includegraphics[width=86mm]{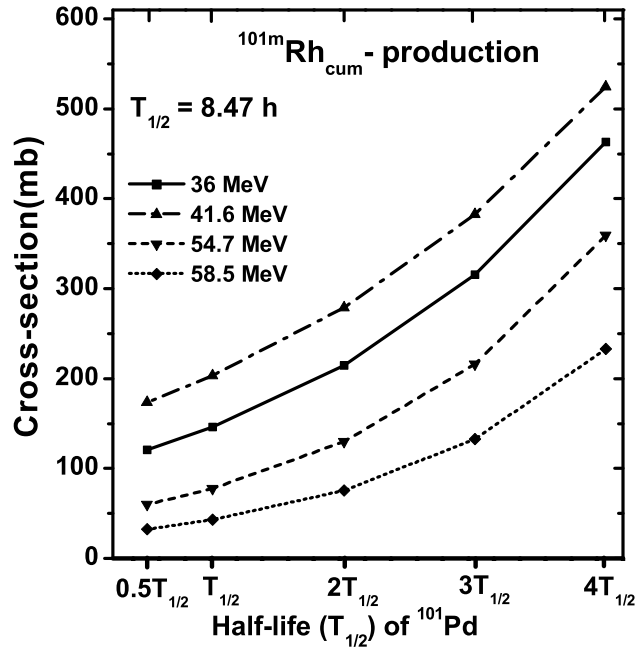}
		\caption{\label{fig14} Cumulative production of $^{101m}$Rh, via its higher charge isobar, $^{101}$Pd (8.47 h) with the passage of time at various incident energies.}
	\end{figure}

\begin{figure} 
		\includegraphics[width=86mm]{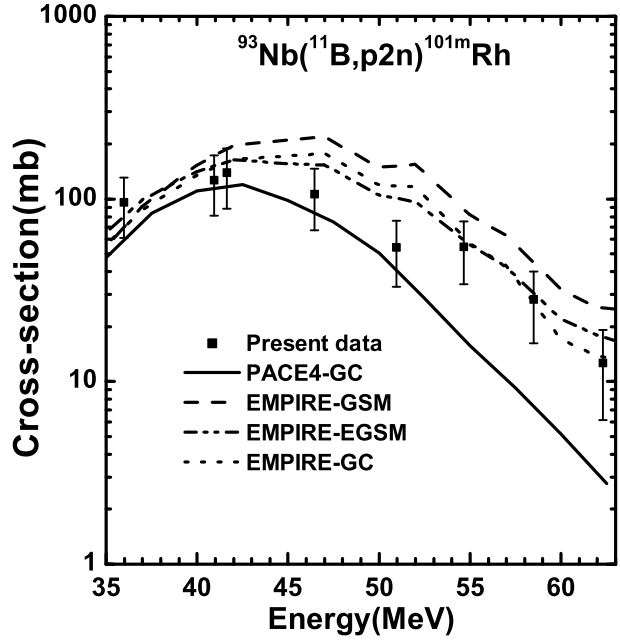}
		\caption{\label{fig15} Same as Fig. \ref{fig11} for $^{101m}$Rh.}
		\end{figure}

\begin{figure}
		\includegraphics[width=86mm]{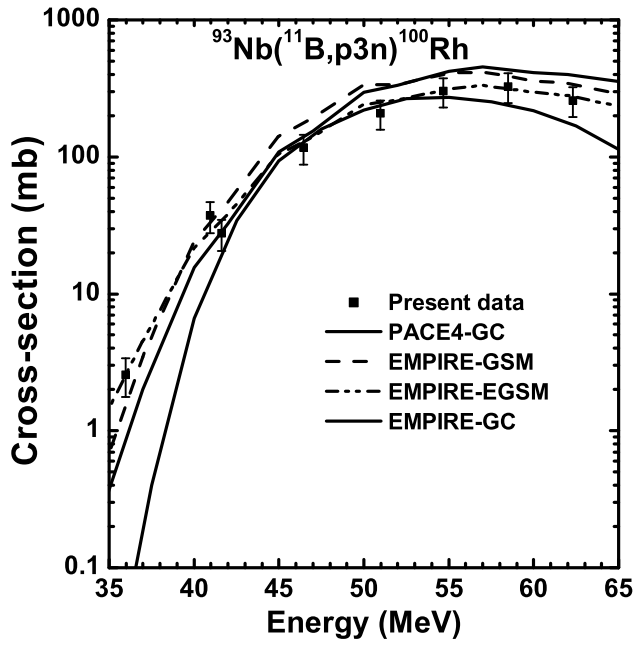}
		\caption{\label{fig16} Same as Fig. \ref{fig11} for $^{100}$Rh.}
	\end{figure}

\begin{figure} 
		\includegraphics[width=86mm]{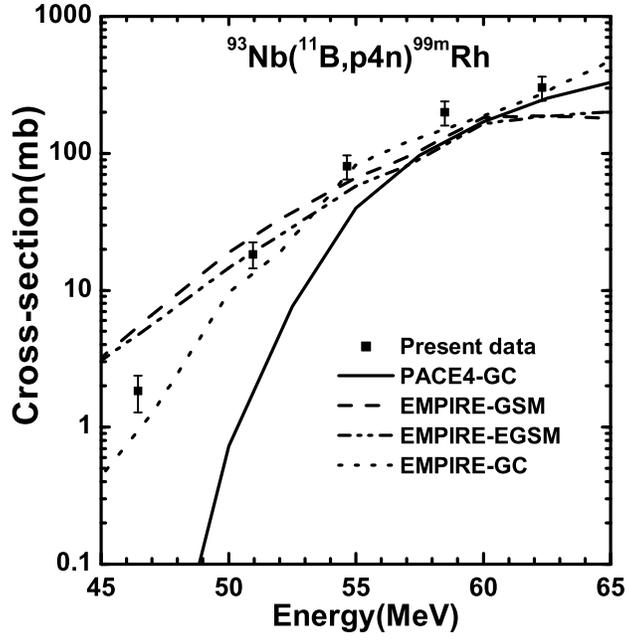}
		\caption{\label{fig17} Same as Fig. \ref{fig11} for $^{99m}$Rh.}
		\end{figure}

\begin{figure}
		\includegraphics[width=86mm]{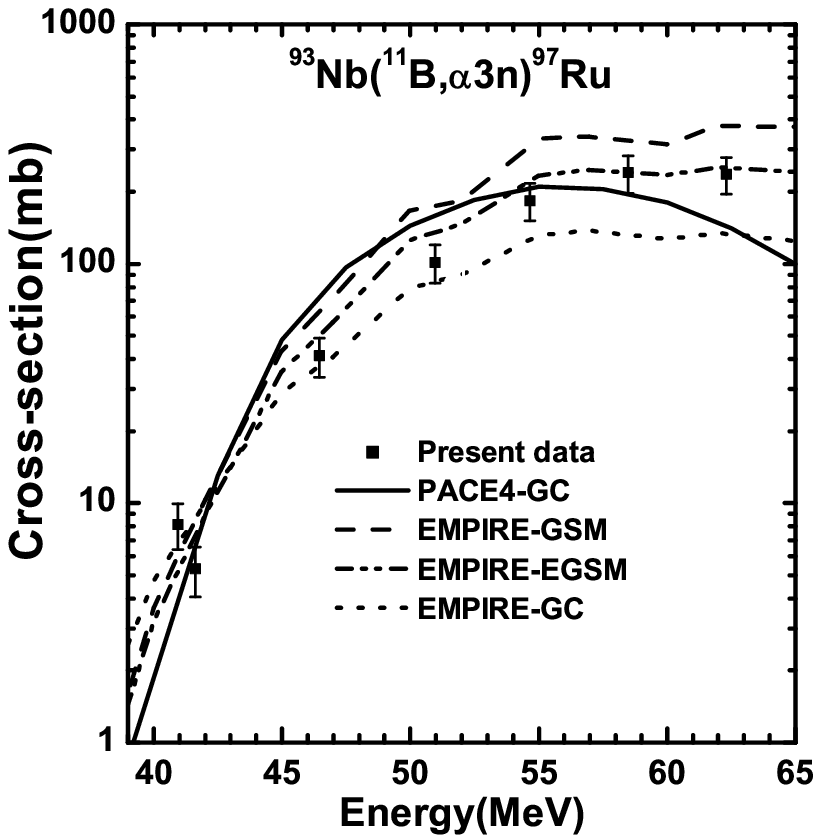}
		\caption{\label{fig18} Same as Fig. \ref{fig11} for $^{97}$Ru.}
	\end{figure}

Production of $^{100}$Pd and $^{99}$Pd at different incident energies are indicated in Fig \ref{fig12} and Fig \ref{fig13}, respectively, to compare with the theoretical results. It is evident from the Fig \ref{fig12} that \textsc{PACE} calculation underpredicts the data at the lower energies and overpredicts at the higher energy range.  Experimental data are well reproduced by the \textsc{EMPIRE} calculation with EGSM level density throughout the energy range within the experimental error. \textsc{EMPIRE} estimation with GC level density overestimates the measured data, while that with the GSM level density well reproduces up to 43 MeV energy and underestimates beyond it. On the other hand, in the Fig \ref{fig13}, none of the model calculation explained the experimental observations, although \textsc{EMPIRE} calculations mimic the trend well.

	\subsubsection{$^{101m,100,99m}$Rh}
	$^{101m}$Rh can be produced in the target matrix either by independent reaction channels or through the decay of its precursor, $^{101}$Pd. Due to the comparatively shorter half-life, $^{101}$Pd (8.47 h) would  decay to $^{101m}$Rh (4.34 d), hence cumulative production would increase if a long cooling time is allowed. Cumulative yield of $^{101m}$Rh in different half-lives of $^{101}$Pd has been shown in Fig \ref{fig14} for various projectile energies. Exponential increment of the cross-section through the decay of its higher charge isobar ($^{101}$Pd) has been observed with the passage of time, as expected. On an average, $\sim$55\% increment in the cross section of $^{101m}$Rh is observed for all the energies at the first half-life of $^{101}$Pd after the EOB. However, independent production cross-sections of $^{101m}$Rh is measured in this work and the excitation function is shown in Fig \ref{fig15}. \textsc{PACE} explains the data up to 51 MeV incident energy and underpredicts above it. \textsc{EMPIRE} calculations with EGSM and GC level density reproduce the experimental data within the error bar over the energy region, while GSM overestimates beyond 40 MeV projectile energy. Nevertheless, \textsc{EMPIRE} calculation shades light on the role of the PEQ reaction mechanism to $^{101m}$Rh at the high energy tail. 
	
	    The $^{100}$Rh has a metastable state $^{100m}$Rh (4.6 m, IT = 98.3 \% and $\epsilon$ = 1.7 \%) and a relatively long lived ground state $^{100}$Rh (20.8 h). Metastable $^{100m}$Rh could not be identified due to its short half-life. Thus measured cross section of $^{100}$Rh is the collective cross section contributed  by the independent production routes and the decay from $^{100m}$Rh, if it was produced. Various possible production routes of $^{100}$Rh  are listed in Table \ref{tab1}. Fig \ref{fig16} shows a comparison between the measured data and the calculations for $^{100}$Rh. \textsc{EMPIRE} calculations satisfactorily predict the experimental results, while \textsc{PACE} calculation shows a small deviation both in lower and higher energy side. \textsc{EMPIRE} with EGSM has best reproduced the experimental results. Although decay of $^{100}$Pd (3.63 d) could add to the cross section of $^{100}$Rh, this possibility was avoided recording  the $\gamma$-ray spectrum  immediately after the EOB. 
		
		The metastable $^{99m}$Rh (4.7 h) that mainly decays to stable $^{99}$Ru ($\epsilon$ $\ge$ 99.84 \%) was produced in the target matrix.  \textsc{EMPIRE} calculation also support the isomeric production along with minute production of ground state $^{99}$Rh. The measured excitation function of $^{99m}$Rh is shown in the Fig \ref{fig17} and compared the isomeric production obtained from \textsc{EMPIRE} along with the \textsc{PACE} that calculates the sum of isomeric and ground state production. \textsc{EMPIRE} with GC explains the experimental result slightly better than EGSM and GSM level density calculation. \textsc{PACE} grossly underestimates the measured excitation function throughout the entire energy range. 
		
	\subsubsection{$^{97}$Ru}
	   Fig \ref{fig18} represents a comparison between experimental data of $^{97}$Ru and the theoretical estimations. Measured results are best described by the \textsc{EMPIRE} with EGSM level density. \textsc{EMPIRE} estimate with GC underpredicts and that with GSM overpredicts the measured excitation function above 52 MeV impinging energy. \textsc{PACE} data satisfactory explain the experimental observation.

			In general, measured cross section data is well reproduced by \textsc{EMPIRE} over \textsc{PACE4}. In the low energy region, a gradual increase in the theoretical cross sections of the radionuclides is observed in EMPIRE calculations compared to the steep increment in \textsc{PACE4}. It might be due to the more accurate treatment of fusion cross section (CCFUS) calculation and a general consideration of Ignatyuk energy-dependent level density parameter in \textsc{EMPIRE}.  Moreover, EMPIRE successfully explained the PEQ emission of particles in both the reactions using exciton model over the compound process. Though the effectiveness of the level density models (EGSM/GSM/GC) could not be concluded precisely from this study, yet GSM/EGSM explains satisfactory most of the reaction channels.

	\subsection{Medically relevant radionuclides} 	
	Significant cross section (above 200 mb) of $^{97}$Ru radionuclide is observed in the 34.6--47.1 MeV wide energy spectrum in which 40.7 MeV projectile energy witnessed the maximum production cross section of $\sim$ 389 mb. Production of most of the contaminants such as $^{95}$Ru, $^{94}$Tc, $^{93m}$Mo and $^{90m}$Y are negligibly small in the desired energy range and would ultimately decay out due to their short half-lives. Therefore, only $^{97}$Ru and $^{95}$Tc will be abundant after a day, which can be separated chemically \cite{maiti13,maiti15,dkumar17}.
	
	On the other hand, production of $^{101m}$Rh (direct and cumulative ) is possible within 36--51 MeV energy range in the $^{11}$B+$^{93}$Nb reaction. Maximum production cross sections of $^{101m}$Rh (139 mb) and its precursor $^{101}$Pd (229.7 mb) that decays into $^{101m}$Rh is measured in the 41.6 MeV energy. A significant quantity of $^{101m}$Rh could be achieved after an appropriate cooling time. 
	
	It is true that heavy ion reactions cannot compete with the light ion production routes (of $^{97}$Ru, $^{101m}$) in terms of the production yield. However, sufficient production of those radionuclides is possible to conduct the  experiments in the laboratory scale. A good quantity of those radionuclides could be produced using a high current heavy ion accelerator.The significance of the heavy ion induced production route of $^{97}$Ru and $^{101m}$Rh from the target matrices Y and Nb, respectively, and their subsequent chemical separation procedures have been discussed in detail in our recent articles \cite{dkumar17, dm17}.
	
		
   	\section{\label{s5}Summary}
    For the first time, a detailed investigation of the excitation functions of the evaporation residues produced in the $^{11}$B-induced reaction on $^{89}$Y and $^{93}$Nb targets were carried out and the theoretical support was provided in order to extract information about their production mechanism. It might be helpful  to optimize the production parameters of medically relevant  radionuclides: $^{97}$Ru, $^{101m}$Rh. 	
		A comparative study of the various level density models (GSM/EGSM/GC) used in \textsc{EMPIRE} is reported in this article. Although, no specific level density model could be suggested to explain the $^{11}$B+$^{89}$Y and $^{11}$B+$^{93}$Nb reaction data, EGSM/GSM calculations reproduced the measured data in many channels.  In general, EMPIRE calculations show better reproduction of the cross sections in most of the channels. The comparisons of the theoretical model calculations suggest that the compound reaction mechanism is the dominant route for the production of residues.
		Like other heavy-ion induced reactions, PEQ emission of neutrons was observed in the $3n$ channels for both the reactions. A signature of PEQ emission was also evident in the $p2n$ and $p3n$ channel in $^{11}$B+$^{93}$Nb reaction. 
Unlike theoretical expectation, a steady and substantial production of $^{90m}$Y radionuclides was observed at all energies which might be the consequence of the neutron transfer processes in $^{11}$B+$^{89}$Y reaction. Satisfactory reproduction of measured data by the EMPIRE calculations indicate the dependence of collective effects, considered in EGSM/GSM level density, on the compound and PEQ processes and the effectiveness of CCFUS calculations used for heavy ion fusion cross section.


	\begin{acknowledgments}
	Authors sincerely acknowledge the technical help and cooperation received from Professor S. Lahiri, SINP, Kolkata, during the experiment, and also thank the Pelletron staff, target laboratory staff of the BARC-TIFR Pelletron facility for their assistance during the experiment. Fellowship from MHRD; grants of IITR/SRIC/218/FIG, Government of India, are gratefully acknowledged.
	\end{acknowledgments}
	


\begin{table*}
		\caption{Spectoscopic decay data \cite{nndc} for the observed radioisotopes and list of contributing process in the $^{11}$B + $^{89}$Y and $^{11}$B + $^{93}$Nb reactions }
		\label{tab1}
		\begin{ruledtabular}
			\begin{center}
				\renewcommand{\arraystretch}{0.6}
				\begin{tabular}{ccccccc}
					Nuclides(J$^\pi$) & Half-life & decay mode(\%) & E$\gamma$(keV) & I$\gamma$(\%) &Contributing reactions & E$_{th}$~\footnote{E$_{th}$ denotes threshold energy.}(MeV)\\
					\hline
					$^{97}$Ru(5/2$^+$)  & 2.83 d& $\epsilon$(100 )         &  215.70 & 85.62  & $^{89}$Y($^{11}$B, 3n)  & 19.26 \\
					           &       &                          &  324.49 & 10.79  &           \\
					$^{95}$Ru(5/2$^+$)  & 1.64 h& $\epsilon$(100)&  336.40 & 70.20  & $^{89}$Y($^{11}$B, 5n)  & 40.39 \\
					           &       &                          &  626.63 & 17.80  &    &  \\
					$^{96}$Tc(7$^+$)  & 4.28 d& $\epsilon$(100)&  778.22 & 99.76  & $^{89}$Y($^{11}$B, p3n) & 27.78 \\
					           &       &                          &  812.54 & 82.00  & $^{89}$Y($^{11}$B, d2n) & 25.28 \\
					           &       &                          &  849.86 & 98.00  & $^{89}$Y($^{11}$B, tn)& 18.25 \\
					$^{95}$Tc(9/2$^+$) & 20.0 h  & $\epsilon$(100)&  765.79 & 93.80  & $^{89}$Y($^{11}$B, p4n) & 36.63 \\
					           &       &                          &         &        & $^{89}$Y($^{11}$B, d3n) & 34.13 \\
					           &       &                          &         &        & $^{89}$Y($^{11}$B, t2n) & 27.10\\  
					$^{94}$Tc(7$^+$)  & 4.88 h& $\epsilon$(100)   & 702.62  & 99.60  & $^{89}$Y($^{11}$B, p5n) & 47.80\\
				               &       &                          & 871.09 & 99.99   & $^{89}$Y($^{11}$B, d4n)& 45.30 \\
					           &       &                          &         &        & $^{89}$Y($^{11}$B, t3n) & 38.26\\		
					$^{93m}$Mo(21/2$^+$) & 6.85 h& IT~\footnote{IT denotes isomeric transition.}(99.88)                & 263.06 & 56.7    & $^{89}$Y($^{11}$B, $\alpha$3n)& 21.21\\
					           &       &                          & 684.67 & 99.7    & $^{89}$Y($^{11}$B, 2p5n) & 53.01 \\
					$^{90m}$Y(7$^+$)  & 3.19 h& IT(100)                & 202.51 & 97.30   & $^{89}$Y($^{11}$B, 5p5n)  & 77.94\\
					           &       &                          & 479.17 & 90.74   & $^{89}$Y($^{11}$B, $\alpha$3p3n)& 46.14  \\
									   &       &                          &  & & $^{89}$Y($^{11}$B, 2$\alpha$pn)& 14.34 \\
					$^{101}$Pd(5/2$^+$) & 8.47 h& $\epsilon$(100)& 296.29  &19.00   & $^{93}$Nb($^{11}$B, 3n) & 19.38 \\
				               &       &                          & 590.44 & 12.06   &   & \\
					$^{100}$Pd(0$^+$) & 3.63 d& $\epsilon$(100 )         & 84.02 & 52.00    & $^{93}$Nb($^{11}$B, 4n) & 28.64\\
					           &       &                          & 126.05 & 7.80    &  & \\
					$^{99}$Pd(5/2$^+$)  & 21.4 m& $\epsilon$(100)& 136.01 & 73.00   & $^{93}$Nb($^{11}$B, 5n) & 41.07 \\
					           &       & & 263.60 & 15.2 & &  \\
					$^{101m}$Rh(9/2$^+$)& 4.34 d& $\epsilon$(92.80),IT(7.20)&306.86 & 84.00  & $^{93}$Nb($^{11}$B, p2n) & 16.29 \\
					           &       &                           &       &        & $^{93}$Nb($^{11}$B, d2n) & 13.80 \\
					           &       &                           &       &        & $^{93}$Nb($^{11}$B, t)   & 6.80 \\
					$^{100}$Rh(1$^-$) & 20.8 h& $\epsilon$(100) &539.60 & 80.60  & $^{93}$Nb($^{11}$B, p3n) & 27.36 \\
					           &       &                           &822.65 & 21.09  & $^{93}$Nb($^{11}$B, d2n) & 24.87 \\ 
					           &       &                           &822.65 & 21.09  & $^{93}$Nb($^{11}$B, tn) & 17.87 \\            
					$^{99m}$Rh(9/2$^+$)  & 4.7 h & $\epsilon(\ge99.84)$&340.71&70.00& $^{93}$Nb($^{11}$B, p4n) & 36.40 \\
					           &       &                           & 617.80 & 12.00 & $^{93}$Nb($^{11}$B, d3n)& 33.91 \\
					           &       &                           &       &        & $^{93}$Nb($^{11}$B, t2n) & 26.91 \\
					$^{97}$Ru(5/2$^+$)  & 2.83 d& $\epsilon$(100)           & 215.70 & 85.62 & $^{93}$Nb($^{11}$B, 2p5n) & 52.97 \\
					           &       &                           & 324.49 & 10.79 & $^{93}$Nb($^{11}$B, $\alpha$3n)   &21.32 \\
				\end{tabular}
			\end{center}
		\end{ruledtabular}
	\end{table*}
	
	\begin{table*}{}
     	\caption{Cross section (mb) of radioisotopes at various incident energies in the $^{11}$B + $^{89}$Y reaction.}
     	\label{tab2}
     	\begin{ruledtabular}
     		\begin{center}
     			\renewcommand{\arraystretch}{1.5}
     			\begin{tabular}{cccccccc}
     				\multirow{3}{1.5cm}{\bfseries Energy (MeV)} \\
     				& \multicolumn{7}{c} { \bfseries Cross-section (mb)}\\
     				\cline{2-8}
     				& \multicolumn{1}{c}{$^{97}$Ru}
     				& \multicolumn{1}{c}{$^{95}$Ru}
     				& \multicolumn{1}{c}{$^{96}$Tc}
     				& \multicolumn{1}{c}{$^{95}$Tc}
     				& \multicolumn{1}{c}{$^{94}$Tc}
     				& \multicolumn{1}{c}{$^{93m}$Mo}
     				& \multicolumn{1}{c}{$^{90m}$Y} \\			\hline
     				27.9 &  1.1 $\pm$   0.6 &                    &                   &                  &                  &                 &                \\		
     				33.2 &103.6 $\pm$  18.8 &                    &                   &                  &                  &                 &                 \\
     				34.6 & 259.6 $\pm$ 46.1 &                    &                &                  &                  &                 &                 \\
     				37.3 & 353.4 $\pm$ 58.5 &                    & 4  $\pm$ 1.6&                  &                  &                 & 0.3 $\pm$ 0.1 \\
     				40.5 & 364.4 $\pm$ 64.2 &                    & 14.9 $\pm$ 2.1  &  0.3  $\pm$ 0.2&                  & 1.1 $\pm$ 0.2 & 0.5 $\pm$ 0.1 \\
     				40.7 & 389.3 $\pm$ 68.9 &                    & 21.8 $\pm$ 3.2  & 0.6  $\pm$ 0.3 &                  & 2.1 $\pm$ 0.6 & 0.6 $\pm$ 0.2 \\
     				41.3 & 333.2 $\pm$ 55.7 &                    & 29.6 $\pm$ 4.1  & 0.8  $\pm$ 0.4 &                  &3.1  $\pm$ 0.7 & 0.7 $\pm$ 0.2 \\
     				45.9 & 280.1 $\pm$ 49.8 & 0.3 $\pm$  0.2   & 123.9 $\pm$ 16& 3.6  $\pm$ 0.8 &                  &25.9 $\pm$ 2.9 & 2 $\pm$ 0.6 \\
     				47.1 & 239.4 $\pm$ 42.4 &  0.7  $\pm$ 0.2  & 116.8 $\pm$ 14.8& 3.6  $\pm$ 0.8 &                  &26.1 $\pm$ 2.7 & 2.1 $\pm$ 0.5 \\
     				50.8 & 131.7 $\pm$ 23.7 & 14.7  $\pm$  4.6 & 214.3 $\pm$ 27.2& 33.9 $\pm$ 6.2 & 0.2 $\pm$ 0.1  &70.5 $\pm$ 7.5 & 4.2 $\pm$ 1.1 \\
     				53.2 &  93.9 $\pm$ 16.8 & 25.5  $\pm$  5.2 & 212.6 $\pm$ 26.5& 59.6 $\pm$ 10.5& 0.5 $\pm$ 0.2  &79.6 $\pm$ 8.1 & 4.2$\pm$ 0.9  \\
     				58.7 &  41.8 $\pm$ 7.7  &  93.1 $\pm$ 18.7 & 229.8 $\pm$ 28.7& 223.2 $\pm$25.9& 2.8  $\pm$ 0.6 &138.3$\pm$ 13.9& 7.3 $\pm$ 1.5 \\	
     			\end{tabular}
     		\end{center}
     	\end{ruledtabular}
     \end{table*}
     
\begin{table*}
     	\caption{Cross section (mb) of radioisotopes at various incident energies in the $^{11}$B + $^{93}$Nb reaction.}
     	\label{tab3}
     	\begin{ruledtabular}
     		\begin{center}
     			\renewcommand{\arraystretch}{1.5}
     			\begin{tabular}{cccccccc}
     				\multirow{3}{1.5cm}{\bfseries Energy (MeV)} \\
     				& \multicolumn{6}{c} { \bfseries Cross-section (mb)}\\
     				\cline{2-8}
     				& \multicolumn{1}{c}{$^{101}$Pd}
     				& \multicolumn{1}{c}{$^{100}$Pd}
     				& \multicolumn{1}{c}{$^{99}$Pd}
     				& \multicolumn{1}{c}{$^{101m}$Rh}
     				& \multicolumn{1}{c}{$^{100}$Rh}
     				& \multicolumn{1}{c}{$^{99m}$Rh} 
     				& \multicolumn{1}{c}{$^{97}$Ru}        \\			\hline
                 	30.6 & 20.3  $\pm$ 4.1  &                    &                  &   &                 &                 &                  \\	
     			    36 & 187.4 $\pm$34  & 9 $\pm$ 1.9    &                  & 96.1 $\pm$ 35.1& 2.6 $\pm$ 0.8 &                 &                  \\	
     				41 & 225.9 $\pm$ 40.8 & 63.9 $\pm$ 12.3  &                  & 127$\pm$ 45.9& 37.2$\pm$ 9.4 &                 & 8.2  $\pm$ 1.8 \\
     				41.6 & 229.7 $\pm$ 41.8 & 51.3  $\pm$ 10  &                  & 139$\pm$ 50.3& 27.7$\pm$ 7.1 &                 & 5.3  $\pm$ 1.3  \\
     				46.4 & 170.1 $\pm$ 31.5 & 134.5  $\pm$25.6 &                  & 106.6$\pm$ 39.4& 116.6$\pm$28.8& 1.8 $\pm$ 0.6 & 41.3 $\pm$ 7.7  \\
     				51 & 94.2 $\pm$ 18.3  & 177.5 $\pm$ 33.7 & 3.7 $\pm$ 1.4  & 54.4 $\pm$ 21.4& 208.1$\pm$51.1& 18.4 $\pm$ 4& 101.6 $\pm$ 18.5 \\
						54.7 & 63.9 $\pm$ 11.7  & 296.9 $\pm$ 65.4 & 13.3 $\pm$ 5.1& 54.7 $\pm$ 20.6  & 302.8 $\pm$ 73.3& 80.5$\pm$16.1& 183.6 $\pm$ 32.4 \\
						58.5 & 38.3 $\pm$ 7.4  & 260.5 $\pm$ 58.6 & 32.1 $\pm$ 11.6& 28.1 $\pm$ 11.9  & 327 $\pm$ 79.2& 199.8$\pm$40.1& 240.2 $\pm$ 42.5 \\
						62.3 & 21.1 $\pm$ 4.4  & 170.5 $\pm$ 40.1 & 42.5 $\pm$ 15.1& 12.7 $\pm$ 6.5  & 258.4 $\pm$ 62.6& 301.4$\pm$59.7& 236.5 $\pm$ 41.7 \\
     			\end{tabular}
     		\end{center}
     	\end{ruledtabular}
			\end{table*}
		

\begin{thebibliography}{9}
		
\bibitem{back14} B. B. Back, H. Esbensen, C. L. Jiang, K. E. Rehm, Rev. Mod. Phy. 86, 317 (2014).

\bibitem{canto15} L. F. Canto, P. R. S. Gomes, R. Donangelo, J. Lubian, M. S. Hussein, Phys. Rep. 596, 1 (2015).
		
\bibitem{parker91} D. J. Parker, P. Vergani, E. Gadioli, J. J. Hogan, F. Vettore, E. Gadioli-Erba, E. Fabrici, M. Galmarini, Phys. Rev. C 44, 1528 (1991).

\bibitem{kelly00} G. R. Kelly, N. J. Davis, R. P. Ward, B. R. Fulton, G. Tungate, N. Keeley, K. Rusek, Phys. Rev. C 63, 024601 (2000).

\bibitem{dasgupta04} M. Dasgupta, P. R. S. Gomes, D. J. Hinde, S. B. Moraes, R. M. Anjos, A. C. Berriman, R. D. Butt, N. Carlin, J. Lubian, C. R. Morton, J. O. Newton, A. Szanto de Toledo, Phys. Rev. C 70, 024606 (2004).

\bibitem{zolnowski78} D. B. Zolnowski, H. Yamada, S. E. Cala, A. C. Kahler, T. T. Sugihara, Phys. Rev. Lett. 41, 92 (1978).

\bibitem{timmers98} H. Timmers, D. Ackermann, S. Beghini, L. Corradi, J. H. He, G. Montagnoli, F. Scarlassara, A. M. Stefanini, N. Rowley, Nucl. Phys. A 633, 421 (1998).
\bibitem{vergani93} P. Vergani, E. Gadioli, E. Vaciago, E. Fabrici, E. Gadioli Erba, M. Galmarini, G. Ciavola, C. Marchetta, Phys. Rev. C 48, 1815 (1993).

\bibitem{sharma15} M. K. Sharma, P. P. Singh, D. P. Singh, A. Yadav, V. R. Sharma, I. Bala, R. Kumar, Unnati, B. P. Singh, R. Prasad, Phys. Rev. C 91, 014603 (2015).

\bibitem{shrivastava06} A. Shrivastava, A. Navin, N. Keeley, K. Mahata, K. Ramachandran, V. Nanal, V. V. Parkar, A. Chatterjee, S. Kailas, Phys Lett B 633, 463 (2006).

\bibitem{hinde11} D. J. Hinde, R. du Rietz, M. Dasgupta, Eur. Phys. J. Web Conf. 17, 04001 (2011).

\bibitem{hinde08} D. J. Hinde, R. G. Thomas, R. du Rietz, A. Diaz-Torres, M. Dasgupta, M. L. Brown, M. Evers, L. R. Gasques, R. Rafiei, M. D. Rodriguez, Phys. Rev. Lett. 100, 202701 (2008).

\bibitem{wallerstein97} G. Wallerstein $et$ $al.$, Rev. Mod. Phys. 69, 995 (1997).

\bibitem{jiang05} C. L. Jiang, K. E. Rehm, H. Esbensen, R. V. F. Janssens, B. B. Back, C. N. Davids, J. P. Greene, D. J. Henderson, C. J. Lister, R. C. Pardo, T. Pennington, D. Peterson, D. Seweryniak, B. Shumard, S. Sinha, X. D. Tang, I. Tanihata, S. Zhu, P. Collon, S. Kurtz, M. Paul, Phys. Rev. C 71, 044613 (2005).

\bibitem{stefanini08} A. M. Stefanini, G. Montagnoli, R. Silvestri, S. Beghini, L. Corradi, S. Courtin, E. Fioretto, B. Guiot, F. Haas, D. Lebhertz, N. M$\breve{a}$rginean, P. Mason, F. Scarlassara, R.N. Sagaidak, S. Szilner, Phys. Rev. C 78, 044607 (2008).

\bibitem{rehm98} K. E. Rehm, H. Esbensen, C. L. Jiang, B. B. Back, F. Borasi, B. Harss, R. V. F. Janssens, V. Nanal, J. Nolen, R. C. Pardo, M. Paul, P. Reiter, R. E. Segel, A. Sonzogni, J. Uusitalo, and A. H. Wuosmaa, Phy. Rev. Lett. 81, 3341 (1998).

\bibitem{vinod13} A. M. Vinodkumar, W. Loveland, R. Yanez, M. Leonard, L. Yao, P. Bricault, M. Dombsky, P. Kunz, J. Lassen, A. C. Morton, D. Ottewell, D. Preddy, and M. Trinczek, Phys. Rev. C 87, 044603 (2013).

\bibitem{signorini97} C. Signorini, Nucl. Phys. A 616, 262c (1997).

\bibitem{maiti09} M. Maiti, S. Lahiri, Phys. Rev. C 79, 024611 (2009).

\bibitem{srivastava78} S. C. Srivastava, P. Som, G. Meinken, A. Sewatkar, T. H. Ku, Brookhaven National Laboratory Report BNL 24614 (1978).

\bibitem{lagunas83} M. C. Lagunas-Solar, M. J. Avila, N. J. Nvarro, P. C. Johnson,  Int. J. Appl. Radiat. Isot. 34, 915 (1983).

\bibitem{zaitseva92}	N. G. Zaitseva,  E. Rurarz, M. Vobecky, K. H. Hwan, K. Nowak, T. Tethal, V. A. Khalkin, L. M. Popinenkova, Radiochim. Acta 56, 59 (1992). 

\bibitem{comar76}	D. Comar, C. Crouzel, Radiochem. Radioanal. Lett, 27, 307 (1976).
 
\bibitem{omperetto80} G. Omperetto, S. M. Quim, Radiochim. Acta 27, 177 (1980).

\bibitem{pao81} P. J. Pao, J. L. Zhou, D. J. Silvester, S. L. Waters, Radiochem. Radioanal. Lett. 46, 21 (1981).

\bibitem{deepak16} D. Kumar, M. Maiti, S. Lahiri, Phys. Rev. C 94, 044603 (2016).

 \bibitem{maiti11} M. Maiti, S. Lahiri, Radiochim. Acta 99, 359 (2011).
 
 \bibitem{maiti15} M. Maiti, S. Lahiri, Radiochim. Acta 103, 7 (2015).
   
 \bibitem{maiti13} M. Maiti, Radiochim. Acta 101, 437 (2013).
 
 \bibitem{dkumar17} D. Kumar, M. Maiti, and S. Lahiri, Sep. Sci. and Tech. (2017) DOI: 10.1080/01496395.2017.1279179.
 
 \bibitem{scholtz77}	K. L. Scholz, V. J. Sodd, J. W. Blue, Int. J. Appl. Radiat. Isot. 28, 207 (1977).
 
 \bibitem{lagunas82} M. C. Lagunas-Solar, S. R. Wilkins, D. W. Paulson, J. Radioanal. Chem. 68, 245 (1982).
 
 \bibitem{lagunas84} M. C. Lagunas-Solar, M. J. Avila, P. C. Johnson, Int. J. Appl. Radiat. Isot. 35, 743 (1984).
 
\bibitem{ditroi07} F. Ditr$\acute{o}$i, F. T$\acute{a}$rk$\acute{a}$nyi, S. Tak$\acute{a}$cs, I. Mahunka, J. Csikai, A. Harmanne, M. S. Uddin, M. Hagiwara, M. Baba, T. Ido, Yu. Shubin, A. I. Dityuk, J. Radional. Nucl. Chem. 272, 231 (2007).
 
 \bibitem{hermanne00} A. Hermanne, M. Sonck, A. Fenyvesi, L. Daraban, Nucl. Instr. Meth. B 170, 281 (2000).
 
 \bibitem{sudar02} S. Sud$\acute{a}$r, F. Cserp$\acute{a}$k, S. M. Qaim, Appl. Radiat. and Isot. 56, 821 (2002).
 
 \bibitem{uddin05} M. S. Uddin, M. Hagiwara, M. Baba, F. Tarkanyi, F. Ditroi, Appl. Radiat. and Isot. 62, 533 (2005).
 
 \bibitem{hermanne02} A. Harmanne, M. Sonck, S. Tak$\acute{a}$cs, F. T$\acute{a}$rk$\acute{a}$nyi , Yu. Shubin, Nucl. Instum. Meth. Phys. Res. B 187, 3 (2002).

 \bibitem{ditroi11} F. Ditr$\acute{o}$i, F. T$\acute{a}$rk$\acute{a}$nyi, S. Tak$\acute{a}$cs, A. Hermanne, H. Yamazaki, M. Baba, A. Mohammadi, A. V. Ignatyuk, Nucl. Instrum. Meth. Phys. Res. B 269, 1963 (2011).
 
 \bibitem{hermanne15} A. Hermanne, F. T$\acute{a}$rk$\acute{a}$nyi, S. Tak$\acute{a}$cs, F. Ditr$\acute{o}$i, Nucl. Instrum. Meth. in Phy. Res. B 362, 110 (2015).
 
 \bibitem{ditroi12} F. Ditr$\acute{o}$i, F. T$\acute{a}$rk$\acute{a}$nyi, S. Tak$\acute{a}$cs, A. Hermanne, A. V. Ignatyuk, M. Baba, Nucl. Instrum. Meth. Phys. Res. B 270, 61 (2012).
 
 \bibitem{skakun08} Ye. Skakun, S. M. Qaim, Appl. Radiat. and Isot. 66, 653 (2008).
 
 \bibitem{ziegler10} J. F. Ziegler, M. D. Ziegler, J. P. Biersack, Nucl. Instrum. Methods Phys. Res. B 268, 1818 (2010).
  
 \bibitem{ms11} M. Maiti, S. Lahiri, Phys. Rev. C 84, 067601 (2011).
  
 \bibitem{m11} M. Maiti, Phys. Rev. C 84, 044615 (2011).
 
 \bibitem{nndc} http://www.nndc.bnl.gov/nudat2/ (National Nuclear Data Center, Brookhaven National Laboratory).
 
 \bibitem{wilke76} B. Wilke, T. A. Fritz, Nucl. Instrum. Methods 138, 331 (1976).
 
 \bibitem{kemmer80} J. Kemmer, R. Hofmann, Nucl. Instrum. Methods 176, 543 (1980).
 
 \bibitem{gavron80} A. Gavron, Phys. Rev. C 21, 230 (1980).
 
 \bibitem{bass77} R. Bass, Phys. Rev. Lett. 39, 265 (1977).
 
 \bibitem{perey76} C. M. Perey, F. G. Perey, At. Data Nucl. Data Tables 17, 1 (1976).
 
  \bibitem{raynal} J. Raynal, Optical-Model and Coupled-Channel Calculations in Nuclear Physics, International Atomic Energy Agency Report
IAEA-SMR-9/8 (1972).
  
  \bibitem{raynal72} J. Raynal, in Computing as a Language of Physics, ICTP International Seminar Course (IAEA-ICTP, Trieste, 1972), p. 281.
  
  \bibitem{iwamoto82} A. Iwamoto and K. Harada, Phys. Rev. C 26, 1821 (1982).

\bibitem{williams71} F. C. Williams, Nucl. Phys. A 166, 231 (1971).
 
 \bibitem{dasso87} C. H. Dasso, S. Landowne, Comp. Phys. Commun. 46, 187 (1987).
 
 \bibitem{hagino12} K. Hagino and N. Takigawa, Progress of Theoretical Physics 128, 1001 (2012).
 
 \bibitem{christensen76} P. R. Christensen, A. Winther, Phys. Lett. B 65, 19 (1976).

\bibitem{dm17} D. Kumar, M. Maiti, and S. Lahiri, 31$^{st}$ DAE-BRNS Nuclear and Radiochemistry Symposium (NUCAR-2017).
 
\end{thebibliography}
\end{document}